\documentclass[aps,prl,twocolumn,groupedaddress,showpacs]{revtex4}
\usepackage{graphicx}
\usepackage{dcolumn}
\usepackage{bm}
\usepackage{amsmath,amsfonts,amssymb,amsthm,verbatim}
\usepackage[ansinew]{inputenc}
\usepackage[usenames]{pstcol}

\usepackage{color}
\usepackage{epsfig}

\newcommand{\beq}{\begin{equation}}
\newcommand{\beqn}{\begin{equation*}}
\newcommand{\eeq}{\end{equation}}
\newcommand{\eeqn}{\end{equation*}}
\newcommand{\beqa}{\begin{eqnarray}}
\newcommand{\beqan}{\begin{eqnarray*}}
\newcommand{\eeqa}{\end{eqnarray}}
\newcommand{\eeqan}{\end{eqnarray*}}
\newcommand{\bdm}{\begin{displaymath}}
\newcommand{\edm}{\end{displaymath}}

\newcommand{\ba}{\begin{array}}
\newcommand{\ea}{\end{array}}

\newcommand\benu{\begin{enumerate}}
\newcommand\eenu{\end{enumerate}}
\newcommand\bit{\begin{itemize}}
\newcommand\eit{\end{itemize}}

\usepackage{epstopdf}
\usepackage{color}
\usepackage{epsfig}
\usepackage{psfrag}

\def\up{\uparrow}
\def\dn{\downarrow}

\def\dx{\partial_x}

\newcommand{\bsp}{\begin{split}}
\newcommand{\esp}{\end{split}}
\newcommand{\bal}{\begin{align}}
\newcommand{\eal}{\end{align}}
\newcommand{\bml}{\begin{multline}}
\newcommand{\eml}{\end{multline}}

\begin{document}
\title{Edge Dynamics in a Quantum Spin Hall State: \\ Effects from Rashba Spin-Orbit Interaction}
\author{Anders Str\"om$^1$, Henrik Johannesson$^1$, and G. I. Japaridze$^{2,3}$}
\affiliation{$\mbox{}^1$ Department of Physics, University of
Gothenburg, SE 412 96 Gothenburg, Sweden} \affiliation{$\mbox{}^2$
Andronikashvili Institute of Physics, Tamarashvili 6, 0177 Tbilisi,
Georgia} \affiliation{$\mbox{}^3$ Ilia State University,
Cholokashvili Avenue 3-5, 0162 Tbilisi, Georgia}
\begin{abstract}
We analyze the dynamics of the helical edge modes of a quantum spin
Hall state in the presence of  a spatially nonuniform Rashba
spin-orbit (SO) interaction. A randomly fluctuating Rashba SO coupling is found to
open a scattering channel which causes localization of the edge
modes for a weakly screened electron-electron (e-e) interaction. A
periodic modulation of the SO coupling, with a wave number
commensurate with the Fermi momentum, makes the edge insulating
already at intermediate strengths of the e-e interaction. We discuss
implications for experiments on edge state transport in a HgTe
quantum well.
\end{abstract}
\maketitle

The quantum spin Hall (QSH) state is a recently discovered phase of
electronic matter in two dimensions which exhibits no symmetry
breaking \cite{KaneMele,Bernevig}. Being an example of a {\em
topological insulator} \cite{HasanKane},  the QSH state is gapped in
the bulk but exhibits massless edge modes. These modes propagate in
opposite directions for opposite spins, each helical mode having a
time-reversed copy. In the range of electron-electron (e-e)
interactions probed in experiments \cite{Konig}, elastic
backscattering induced by a single (nonmagnetic) impurity or by
quenched disorder is suppressed by time-reversal invariance
\cite{Wu,Xu}. For samples of a length smaller than the spin
dephasing length \cite{Jiang} one thus expects that the edge modes
are protected, resulting in nondissipative transport with a
quantized Landauer conductance  $2e^2/h$. This expectation is
supported by experiments \cite{Konig}, holding promise for
exploiting QSH edges in  future integrated circuit technology as a
means to reduce power dissipation as devices become smaller.

The question whether the QSH edge is stable also against e-e
interactions in the presence of a {\em Rashba spin-orbit} (SO)
{\em interaction} has received less attention. One should here recall
that SO interactions $-$ which allow for spin-flip scattering while
respecting time-reversal invariance $-$ come in different guises. In
the case of a HgTe quantum well $-$ the best studied platform for
realizing a QSH state \cite{Konig, BHZ} $-$ the {\em intrinsic} SO
interaction in the atomic $p$ orbitals crucially contributes to the
very special properties of this material: For a quantum well (QW) thicker than a
critical value it inverts the band structure, pushing the $p$ band
above the $s$ band. A simple model calculation reveals the presence
of a single pair of helical edge bands inside the inverted bulk band
gap, showing that this intrinsic SO interaction is what makes
possible the QSH state \cite{Konig2}. This is to be contrasted with
the {\em Rashba} SO interaction which originates from the
gate-controllable inversion asymmetry of a QW \cite{Winkler}. Its
strength can be huge in the bulk metallic phase of an inverted HgTe
structure, with an effective coupling several times larger than for
most other known semiconductor heterostructures \cite{Gui}. In a QSH
state with an insulating bulk, charge and spin are transported by
the edge modes. With the bands for these modes dispersing all the
way from the bulk valence band to the bulk conduction band
\cite{Konig2}, one expects a sizable Rashba interaction to still
operate at the edge of the sample, making it important to inquire
about its effect on the edge modes.

In this Letter we analyze the dynamics at the QSH edge in the
presence of a Rashba SO interaction. Treating the Rashba interaction
as a perturbation, we find that it has no effect on the propagation
of edge modes when it is spatially uniform. The more realistic case
when the Rashba coupling fluctuates randomly in space
\cite{Sherman,Golub} is different, however. Together with a weakly
screened e-e interaction, the Rashba interaction now opens a
scattering channel for the edge modes which may cause them to
localize. For a sufficiently strong e-e interaction we find that the localization length 
is shorter than the length of a typical sample \cite{Konig}, implying that the
edge modes {\em do} localize, causing the QSH state to collapse.
As the screening in a QW can be controlled by varying the
thickness of the insulating layer between the well and the nearest
metallic gate, our prediction is amenable to a direct experimental
test. Interestingly, by subjecting the edge to a periodically
modulated Rashba interaction, with a wave number commensurate with
the Fermi momentum, we find that the edge becomes insulating already
at intermediate strengths of the e-e interaction. A periodic and
tunable Rashba modulation, produced by a sequence of equally spaced
nanosized electrical gates, may thus be used as a current switch in
a QSH-based spin transistor.

To model the edge dynamics of a QSH state we use two electron fields
$\Psi_{\uparrow}(x)\! =\! \psi_{\uparrow}(x)e^{ik_Fx}$ and
$\Psi_{\downarrow}(x)\! =\! \psi_{\downarrow}(x)e^{-ik_Fx}$, where
$\Psi_{\uparrow}$ [$\Psi_{\downarrow}$] annihilates a clockwise
[counterclockwise] propagating edge electron with spin-up [spin-
down], quantized along the growth direction $\hat{z}$
of the QW. The fields $\psi_{\uparrow}$ and $\psi_{\downarrow}$ are slowly
varying, with the fast spatial variations of the original
electron fields contained in the rapidly oscillating $e^{\pm ik_F
x}$ factors. The low-energy kinetics of the helical edge electrons
is then encoded by the one-dimensional Dirac Hamiltonian
\begin{equation} \label{kinetic}
H_0 = -iv_F  \int dx \left(\psi^{\dagger}_{\uparrow}
\partial_x\psi^{\phantom{\dagger}}_{\uparrow} -
\psi^{\dagger}_{\downarrow}\partial_x\psi^{\phantom{\dagger}}_{\downarrow}\right),
\end{equation}
with $v_F$ the Fermi velocity. Away from half filling of the 1D band
of edge states, time-reversal invariance constrains the possible
e-e interaction processes to
dispersive $(\sim g_d)$ and forward $(\sim g_f)$ scattering
\cite{Wu,Xu}. Close to the Fermi points  $\pm k_F$ these
interactions are given by
\begin{equation} \label{dispersive}
H_d \!=\! g_d\!\int dx \,\psi^{\dagger}_{\uparrow}
\psi^{\phantom{\dagger}}_{ \uparrow} \psi^{\dagger}_{\downarrow}
\psi^{\phantom{\dagger}}_{\downarrow}, \ \  H_f \!= \!\frac{g_f}{2}  \!\int dx \, \psi^{\dagger}_{\alpha}
\psi^{\phantom{\dagger}}_{\alpha} \psi^{\dagger}_{\alpha}
\psi^{\phantom{\dagger}}_{\alpha},
\end{equation}
respectively, with repeated spin indices
$\alpha$ summed over. While the forward scattering only leads to
a trivial velocity shift, the dispersive
scattering, controlled by $H_d$, is an
important feature of a QSH edge \cite{SJ}. In the present problem it will be
shown to have dramatic effects.

We now add a Rashba SO
interaction \cite{Winkler},
\beq \label{rashba} H_R= \int  dx\, \alpha(x)  \Psi_\alpha^\dag(x)
\sigma_{\alpha\beta}^y \,k_x \Psi^{\phantom{\dagger}}_{\beta}(x) +
 \mathrm{H.c.}, \eeq
where $\alpha(x)$ is a spatially varying coupling, $\sigma^y$ is a
Pauli matrix, and $k_x$ is the wave number along the edge. Rewriting $H_R$ in terms
of the slowly varying fields $\psi_{\uparrow}$ and
$\psi_{\downarrow}$, with $k_x \!=\! -i\partial_x$, one obtains
\begin{equation} \label{cont}
H_R \!= \!\int \!dx \, \alpha(x)\!
\left((\dx\psi^\dag_{\up})\psi^{\phantom{\dagger}}_{\dn}\! -\!
\psi^\dag_{\up}\dx\psi^{\phantom{\dagger}}_{\dn}\right)e^{-2ik_Fx} \!+ \! \mathrm{H.c.}
\end{equation}

With an eye towards QSH physics in a HgTe QW, a few comments are in
order. First, the lowest bulk conduction band ($H_1$) in an inverted
HgTe structure is nonparabolic, implying that the 
spatially averaged {\em effective} bulk Rashba coupling, $\alpha^{\ast}$ call it,
is $k$ dependent \cite{Konig3}. As an estimate of the spatially
averaged Rashba coupling $\langle \alpha(x)\rangle$ for the edge
states in (\ref{cont}) we shall take the size of $\alpha^{\ast}$ in
the range corresponding to the wave numbers where the two edge bands
join the $H_1$ band.
Second, one should realize that the 
coupling $\alpha(x)$ depends on several distinct features
of the QW, most importantly the applied gate electric field, the ion
distribution in the nearby doping layers \cite{Sherman}, and  the
presence of random bonds at the two QW interfaces \cite{Golub}. As we shall
need no details, we here treat $\alpha(x)$ as a phenomenological
parameter, and write it as $\alpha(x) = \langle \alpha(x) \rangle +
\sum_n \hat{\alpha}(k_n) e^{ik_nx}$, with $\hat{\alpha}(k_n)$ the
Fourier modes of the zero-mean random contribution from the dopant
ions and interface bonds, here taken to obey Gaussian statistics
\cite{Efros}.

With these preliminaries, let us now explore the effect of the
Rashba coupling on the edge dynamics. As expected, the
spatial average $ \langle \alpha(x) \rangle$ is seen to leave no
trace in $H_R$ as the corresponding terms in the integrand oscillate
rapidly and average out upon integration. Hence, to lowest order,
a perturbation with a uniform Rashba interaction has no influence on the low-energy
dynamics of the edge states \cite{FootNote0}.  In fact, the
invariance of $H_R$ under time reversal ${\cal T}$, together with
the ${\cal T}^2 = -1$ property of a single-electron state, implies that the lowest-order effect produced by
$H_R$ for {\em any} $\alpha(x)$ can at most be of  ${\cal O}(\hat{\alpha}^2)$. For
noninteracting electrons, power counting on Eq.\ (\ref{cont}) shows
that an ${\cal O}(\hat{\alpha}^2)$ process is  irrelevant [in renormalization-group 
(RG) sense], implying robustness of the
edge states against perturbations with a Rashba interaction, even
when spatially fluctuating.

To find out how this picture may change in the presence of e-e
interactions, we bosonize the Hamiltonian $H\!=\!H_0 \!+\! H_d\! + H_f
\!+\!H_{R}$, defined by Eqs.\ (\ref{kinetic}), (\ref{dispersive})
and (\ref{cont}), and put $\psi_{\uparrow} =\eta_{\uparrow} \exp
\left({\it i} \sqrt{\pi}[\phi+\theta]\right)/\sqrt{2\pi \kappa}$ and
$\psi_{\downarrow}= \eta_{\downarrow} \exp \left( -{\it
i}\sqrt{\pi}[\phi-\theta]\right) /\sqrt{2\pi \kappa}\label{L}$
 \cite{Giamarchi}. Here $\phi$ and $\theta$ are dual bosonic fields
satisfying $\partial_t \phi = v \partial_x \theta$ with 
$v \!=\![(v_{\small F} \!+ \!g_f/\pi)^2\!-\!(g_d/\pi)^2]^{1/2},$  and
$\eta_{\uparrow \downarrow}$ are Klein factors satisfying 
$\{\eta_{\uparrow}, \eta_{\downarrow}\}\!=\!0$.
The microscopic cutoff $\kappa$ is given by the penetration depth of the edge
states, $\kappa \approx v_F/W$, with $W$ the bulk band gap (in units
with $\hbar \!=\!1$).
Absorbing a factor of $i$ in $\eta_{\downarrow}$, one obtains
\begin{multline}\label{H_Bosonized}
H_0 + H_d + H_f = v \int \!dx \! \left( \frac{1}{2K}(\partial_x\phi)^2 +
\frac{K}{2}(\partial_x \theta)^2 \right)\!,\\
H_R  = \frac{1}{\sqrt{\pi}\kappa} \int dx\, \eta(x)(\dx \theta)
e^{i\sqrt{4\pi}\phi} + \mathrm{H.c.},
\end{multline}
with $\eta(x)\equiv\sum_n \hat{\alpha}(k_n-2k_F)
e^{ik_nx}$, and $K
=[(\pi v_{\small F}+g_f -g_d)/(\pi v_{\small F} +g_f +g_d)]^{1/2}$.

Having obtained the theory on bosonized form, Eq.\ (\ref{H_Bosonized}), we pass
to a Lagrangian formalism by a Legendre transform of the Hamiltonian. We use that
$\Pi = \sqrt{K}\partial_x \theta$ serves as conjugate momentum to $\phi/\sqrt{K}$ and
integrate out $\Pi$ from the partition function $Z$ to arrive at
\begin{equation}
Z \sim  \int {\cal D}\varphi e^{-S[\varphi]},
\end{equation}
with the Euclidean action
\begin{multline} \label{Action}
S[\varphi] = \frac{1}{2}\!\int dx d\tau \Big(\frac{1}{v}(\partial_{\tau}
\varphi)^2 +
v(\partial_x \varphi)^2\Big) \\
- \frac{1}{2\pi \kappa}\int dx d\tau \left(\xi(x) e^{-i \lambda_K \varphi} + \mathrm{H.c.}\right).
\end{multline}
Here $\xi(x) \equiv 1/(4Kv\kappa)\sum_{n,n'}\hat\alpha(k_n-2k_F)\hat\alpha(k_{n'}-2k_F)\times e^{i(k_n+k_{n'})x}$, $\lambda_K \equiv \sqrt{16\pi K}$, and
$\varphi\! \equiv\! \phi/\sqrt{K}$. Note that by integrating out
$\Pi$, terms linear in $\eta$ become proportional to a total time
derivative and hence vanish. Terms proportional to $\eta\,
\eta^{\ast}$ contribute only an immaterial constant.

By averaging over the randomness in $S[\varphi]$, using the Gaussian
statistics $P[\xi(x)] = \mbox{exp}[-D^{-1}_{\xi} \int dx\,
\xi^{\ast}(x) \xi(x)]$ so that $\langle \xi(x) \rangle =0$ and $
\langle \xi^{\ast}(x) \xi(x') \rangle = D_{\xi} \delta(x-x')$, the
replica method \cite{Giamarchi} yields the disorder-averaged action
\begin{multline} 
S_n[\varphi] =  \frac{1}{2}\sum_a\!\int dx d\tau \Big(\frac{1}{v}(\partial_{\tau}
\varphi_a)^2+ v(\partial_x \varphi_a)^2\Big) \\
 -\frac{D_{\xi}}{(2\pi\kappa)^2}\! \sum_{a, b}\!\int \!dx  d\tau d\tau' \!\
cos[\lambda_K \Phi_{ab}(x,\!\tau,\!\tau')] \nonumber \end{multline}
where $a,b\! =\! 1,...,n$ are replica indices with $\Phi_{ab}(x,\!\tau,\!\tau') \equiv \phi_a(x,\tau) - \phi_b(x,\tau')$, and $D_{\xi} =
n_i/(8 \pi K^2 v^4)\left<\mathrm{Re}\,\hat{\alpha}^4(k)\right> $ with $n_i$ the
composite density of the dopant ions and interface bonds that produce the
randomness in the Rashba coupling \cite{Sherman}. The second-order
RG equations of $D_{\xi}, v,$ and $K$, generated by the scaling
$(\tau,x) \rightarrow
(\tau,x)\exp(-\ell)$ ($\ell\!>\!0$), are given by $\partial D_{\xi}/\partial \ell =
(3-8K)D_\xi, \partial v/\partial \ell = -2vK D_\xi$ and $\partial K/\partial
\ell = -2K^2 D_\xi$. It follows that the Rashba coupling grows under
renormalization when $K < K_c = 3/8$,  driving a transition to an Anderson-type localized
state. The value $K_c=3/8$ was identified in Refs.\ \cite{Wu,Xu} as
the critical value below which correlated backscattering in the
presence of quenched disorder may cause localization of the helical edge
modes. Our analysis shows that a Gaussian distributed random Rashba coupling yields a precise microscopic realization of this type of process.

Importantly, to find out whether the edge electrons of a given experimental
sample do become localized, one must test for the condition
$\xi_{\mbox{\footnotesize{loc}}} < L$, with $\xi_{\mbox{\footnotesize{loc}}}$
the localization length and $L$ the length of the sample. 
For this purpose we rewrite the replica RG equations for $D_{\xi}$ and $K$ on the form of Kosterlitz-Thouless (KT) equations, $\partial z_\parallel/\partial \ell=-z_\perp^2$ and $\partial z_\perp/\partial \ell=-z_\parallel z_\perp$, with $z_\parallel\equiv 4K-3/2$ and $z_\perp\equiv 8K\sqrt{D_\xi}$. We renormalize $z_\perp(\ell)$ until $\ell=\ell^{\ast}$, defined by $D_\xi(\ell^{\ast})\sim 1$. The  renormalized localization length $\xi_\mathrm{loc}(\ell^{\ast})$ is of the same size as the cutoff $\kappa$, and since all lengths in the system scale with a factor $\exp(-\ell)$, the true localization length of the system is given by $\xi_\mathrm{loc} \approx \kappa \exp(\ell^{\ast})\, $\cite{GiamarchiSchulz}.

\begin{figure}[tbh]
\centering
\includegraphics[width=\columnwidth]{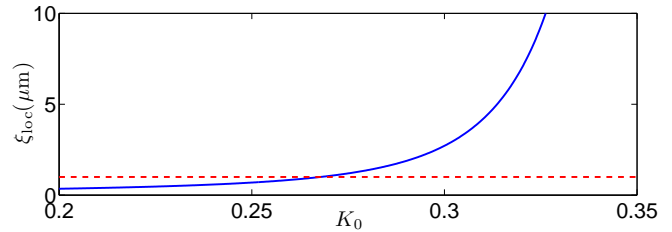}
\centering
\caption{(Color online) The edge localization length $\xi_\mathrm{loc}$ for different values of the interaction parameter $K_0$. The dashed line marks the length of a micron-sized HgTe QW sample \cite{Konig2}.
}\label{llocplot}
\end{figure}

To make an estimate of $\xi_\textrm{loc}$ for a HgTe QW we must put numbers on our parameters. Starting with the random Rashba coupling $\alpha_{\mbox{\footnotesize{rand}}}(x)$, we have that 
$\sqrt{\langle \alpha^2_{\mbox{\footnotesize{rand}}}(x) \rangle} \approx \langle \alpha(x) \rangle$ for a QW with a zinc-blende lattice structure \cite{Sherman}. This, together with the estimate 
$\hbar \langle \alpha(x)\rangle \approx \hbar \alpha^{\ast} \approx 5 \times 10^{-11}$ eVm at the wave numbers at which the edge bands join the $H_1$ band \cite{Konig3}, suggests that $ D_\xi \approx  1.5\times
10^{-3}$, keeping in mind that a normal distributed $\alpha(x)$ yields a normal distributed $\hat\alpha(k)$ with $\left<\mathrm{Re}\,\hat\alpha^2(k)\right>
=\left<\alpha^2(x)\right>/2$, and that $\left<\hat{\alpha}^4\right>\!=\!3\left<\hat{\alpha}^2\right>^2$.
As for the unrenormalized interaction parameter
$K$ (from now on denoted $K_0$), we use that $K_0\!\approx \!(1+\lambda \ln (d/s))^{-1/2}$ for $g_d\! =\! g_f \!\approx \!V(q\!=\!0)/\hbar v_F$ with $V(q)$ the Fourier transform of the Coulomb potential, $d$ 
being the distance to the nearest metallic gate, $s \!= \!\mbox{max}\,[\, \kappa,
d_{\mbox{\footnotesize{QW}}}\,]$ with
$d_{\mbox{\footnotesize{QW}}}$ the thickness of the QW, and $\lambda = 2e^2/(\pi^2
\epsilon \epsilon_0 \hbar v_F)$ with $\epsilon$ the
permittivity of the doping and insulating layers between the gate and the QW. 
By varying the composition and the thickness of the layers, $K_0$ can take values
from $\approx$ 1 (fully screened electrons)  down to  $\sim$ 0.1 (strongly interacting limit), with $0.5 \lesssim \!K_0 \!\lesssim 0.55$ in the experiments reported in Ref.\ \onlinecite{Konig} \cite{FOOTNOTE}. Taking $s \!\approx \!10$ nm, $n_i \approx 10^9/$m, and $v \approx v_F \approx 5 \times 10^5$ m/s \cite{Konig2}, we have all the data needed for estimating how
$\xi_{\mathrm{loc}}$ depends on $K_0$. The result is plotted in Fig.\ \ref{llocplot}, revealing that the edge of a micron-sized sample localizes for $K_0 \lesssim 0.25$.

 \normalsize
\normalsize

When $\xi_\textrm{loc} < L$, the edge electrons are localized up to 
temperatures of the order of $\hbar v/(k_B
\xi_{\mbox{\footnotesize{loc}}})$, with an exponential decrease of
the conductivity for lower temperatures
\cite{GiamarchiSchulz}.  Thus, the edge remains insulating at the
low temperatures (30mK$-$2K) at which experiments are typically
carried out  \cite{Konig}. 

Having explored the Rashba interaction with a spatially
random coupling, one may ask how the QSH edge modes respond to a
periodically modulated coupling. The question
becomes particularly interesting in light of a proposal by Wang
\cite{Wang} to use a gate-controlled Rashba
coupling as a current switch in a quantum wire. As shown in Ref.\
\onlinecite{Wang}, a periodic modulation of a Rashba interaction,
produced by a sequence of equally spaced nanosized gates,
makes the electrons backscatter coherently and the current gets
blocked. By decharging the gates, the current is free to flow again. If
this blueprint for a spin transistor can be copied over to the edge
of a QSH insulator, one may envision a device enabling fast and
efficient control of a dissipationless current.

\begin{figure}[tbh]
\centering
\includegraphics[width=\columnwidth]{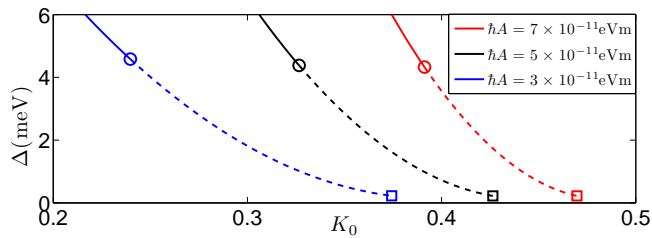}
\centering
\caption{(Color online) The gap $\Delta$ for different Rashba amplitudes $\hbar A$ and values of $K_0$ 
The circles and squares mark the smallest gaps for HgTe QW samples of length $1\,\mu$m and $20\,\mu$m, respectively.
}\label{gapplot}
\end{figure}

As a simple model for a periodically modulated Rashba coupling we
take $\alpha(x) = A \cos(Qx)$ in Eq.\ (\ref{cont}) \cite{Japaridze}.  By
inspection, the largest effect from this modulation is obtained by
choosing $Q = 2k_F$, as this will produce terms in the integrand
where {\em all} oscillations are carried by the slowly varying
fields $\psi_{\uparrow, \downarrow}$ (with $2k_F$ corresponding to a
wavelength of roughly $5 - 10$ nm in a HgTe QW \cite{Konig3}). 
By considering $K_0$ values corresponding
to $\xi_{\mbox{\footnotesize{loc}}} > L$ (cf.\ Fig.\ 1), we can neglect the intrinsic
disordering of the Rashba coupling. Repeating the
procedure that took us from Eqs.\
(\ref{kinetic}), (\ref{dispersive}), and (\ref{cont}) to the action
in Eq.\ (\ref{Action}), but now with $\eta(x)$ replaced by
the amplitude $A$, we obtain
\begin{equation} 
S[\varphi] = \frac{1}{2}\!\int dx d\tau \Big(\frac{1}{v}(\partial_{\tau}
\varphi)^2 + v(\partial_x \varphi) 
- 2g \cos(\lambda_K\varphi)\Big), \nonumber
\end{equation}
with $g\! \equiv \! A^2/(4\pi K \kappa^2v)$. This action is
that of the well-known quantum sine-Gordon model. The RG flows of
$g$ and $K$ are governed by the KT equations
$\partial z_{\parallel}/\partial \ell = -
z_{\perp}^2$ and $\partial z_{\perp}/\partial \ell = - z_{\parallel}z_{\perp}$,
with $z_{\parallel} \equiv 4K-2$ and $z_{\perp} \equiv 4g\sqrt{CK^3}$,
where $C$ is a $\kappa$-dependent constant \cite{Giamarchi}. When $K_0<1/2$, the mass
term in the sine-Gordon action is seen to be RG relevant. In this
regime backscattering generates a
dynamical mass for the edge modes, turning the edge to a Mott
insulator. Whether the corresponding mass gap $\Delta$, measured from the
Fermi level, is large enough to block the current depends on
its size relative to the available thermal energy. To make an
estimate for the case of a HgTe QW of length $1\,\mu$m [$20\,\mu$m], we calculate $\Delta$ for three different values of the Rashba amplitude, tuning up the interaction parameter $K_0$ until the renormalization length $\kappa \exp(\ell^*)$ reaches $1\,\mu$m [$20\,\mu$m], with $\exp(\ell^*)$ the scale factor for which backscattering processes start to dominate (see Fig.\ \ref{gapplot}). When $\kappa \exp(\ell^*)\approx  1\,\mu$m, $\Delta \approx 4.5\,$meV for all three amplitudes, corresponding to a temperature of around $50$K.

\normalsize
\normalsize


The thermally activated conductance is given by $G = (2
e^2/h)e^{-\Delta/k_BT}$. To prevent thermal leakage of current
through the proposed switch, one may thus need to cool a micron-sized sample to
temperatures in the range $10-20$ K. For these temperatures our result provides a
proof-of-concept of using an ''on-off'' modulated Rashba coupling as
a switch for QSH edge currents. It should be interesting to repeat our
analysis for QWs based on the ternary Heusler compounds
\cite{Chadov}, once these have been put together and characterized.
As discussed in Ref.\ \onlinecite{Chadov}, the diversity of Heusler
materials opens wide possibilities for tuning the bulk band gap and
setting the desired band inversion. This may allow also for enhanced
Rashba interactions at the edge. The prospect for a 
QSH-based spin transistor makes this a
promising path to explore.

In summary, we have carried out an analysis of the combined effect of
Rashba spin-orbit and e-e interactions on the edge dynamics in a QSH state. 
The spatial disordering of the Rashba coupling, intrinsic to a quantum well structure that supports a QSH state, is found to localize
 the edge electrons for sufficiently strong e-e interactions. Likewise, a periodic modulation of the coupling $-$ realizable by placing a configuration of equally spaced nanosized gates on top of the quantum well $-$ localizes the electrons already at intermediate strengths of the e-e interaction. While a practical implementation may have to await advances in nanoscale technology, our result suggests that a Rashba-controlled switch for the dissipationless edge current in a QSH device is a viable possibility.  On the more fundamental side, trying to characterize the transition from the Mott insulating phase caused by a periodic Rashba modulation to the Anderson-type localization due to the interplay of e-e interactions and the disordered Rashba coupling remains a challenge.

We thank H. Buhmann, X. Dai, and L. Ioffe for valuable discussions.
HJ acknowledges the hospitality of LPTMS, Universit\'e de Paris-Sud,
for hospitality during the completion of this work. This research
was supported by the Swedish Research Council under Grant No.
621-2008-4358 and the Georgian NSF
Grant No. ST/09-447.


\begin{thebibliography}{99}

\bibitem{KaneMele} C. L. Kane and E.~J.~Mele, Phys. Rev. Lett. {\bf 95}, 226801
(2005).

\bibitem{Bernevig} B. A. Bernevig and S.-C. , Phys. Rev. Lett. {\bf 96},
106802 (2006).

\bibitem{HasanKane}For a review, see M. Z. Hasan and C. L. Kane, arXiv:1002.3895.

\bibitem{Konig} M. K\"onig {\em et al.}, Science {\bf 318}, 766 (2007); A. Roth {\em et al.}, Science {\bf 325}, 294 (2009).

\bibitem{Wu} C. Wu {\em et al.}, Phys. Rev. Lett. {\bf 96}, 106401 (2006).

\bibitem{Xu} C. Xu and J. E. Moore, Phys. Rev. B {\bf 73}, 045322 (2006).

\bibitem{Jiang}H. Jiang {\em et al.}, Phys. Rev. Lett. {\bf 103}, 036803 (2009).

\bibitem{BHZ} B. A. Bernevig {\em et al.}, Science {\bf 314}, 1757 (2006).

\bibitem{Konig2} M. K\"onig {\em et al.},  J. Phys. Soc. Japan {\bf 77}, 031007 (2008).

\bibitem{Winkler}
R. Winkler, {\em Spin-Orbit Interaction Effects in Two-Dimensional Electron and Hole Systems} (Springer
Verlag, Berlin Heidelberg, 2003).

\bibitem{Gui}
Y. S. Gui {\em et al.}, Phys. Rev. B {\bf 70}, 115328 (2004).

\bibitem{Sherman} E. Ya. Sherman, Phys. Rev. B {\bf 67}, 161303(R) (2003).

\bibitem{Golub} L. E. Golub and E. L. Ivchenko, Phys. Rev. B {\bf 69}, 115333 (2004).

\bibitem{SJ}A. Str\"om and H. Johannesson, Phys. Rev. Lett. {\bf 102}, 096806
(2009).

\bibitem{Konig3}M. K\"onig {\em et al.}, Phys. Status Solidi (c) {\bf 4}, 3374 (2007).


\bibitem{Efros} A. L. Efros and B. I. Shklovskii, {\em Electronic Properties of
Doped Semiconductors} (Springer, 1989).

\bibitem{FootNote0}
The known {\em second-order} spin splitting from a uniform Rashba
interaction \cite{Wu} shows up as a constant shift of the spectrum
of the spin-mixed edge modes when treated as a perturbation, with no
effect on their propagation.


\bibitem{Giamarchi}
T. Giamarchi, {\em Quantum Physics in One Dimension}, (Oxford University Press, Oxford, 2003).

\bibitem{GiamarchiSchulz}
T. Giamarchi and H. J. Schulz, Phys. Rev. B {\bf 37}, 325 (1988).

\bibitem{FOOTNOTE}
This estimate of $K_0$ agrees with Ref. \onlinecite{Chamon}. The larger value in Ref.
\onlinecite{SJ} stems from an overestimate of screening.

\bibitem{Chamon}
C.-Y. Hou {\em et al.}, Phys. Rev. Lett. {\bf 102},
076602 (2009).

\bibitem{Wang}
X. F. Wang, Phys. Rev. B {\bf 69}, 035302 (2004).

\bibitem{Japaridze}
G. I. Japaridze {\em et al.}, Phys. Rev. B {\bf 80}, 041308(R) (2009).


\bibitem{Chadov} S. Chadov {\em et al.}, arXiv:1003.0193.

\end{thebibliography}
\end{document}